\documentclass[twocolumn,prl,tightenlines,superscriptaddress,showpacs]{revtex4-2}

\usepackage{amsmath}
\usepackage{amssymb,amsfonts,latexsym}
\usepackage{bm}
\usepackage[mathcal]{euscript}
\usepackage{graphicx}
\usepackage{epsfig}
\usepackage{color}

\newcommand{\be}{\begin{equation}}
\newcommand{\ee}{\end{equation}}
\newcommand{\ve}{\varepsilon}
\newcommand{\la}{\langle}
\newcommand{\ra}{\rangle}
\newcommand{\mC}{\mathcal{C}}

\begin{document}

\title{Nonequilibrium Phase Transition To Temporal Oscillations In Mean-Field Spin Models}

\author{Laura Guislain}
\affiliation{Univ.~Grenoble Alpes, CNRS, LIPhy, 38000 Grenoble, France}

\author{Eric Bertin}
\affiliation{Univ.~Grenoble Alpes, CNRS, LIPhy, 38000 Grenoble, France}

\date{\today}
%\pacs{}

\begin{abstract}
We propose a mean-field theory to describe the nonequilibrium phase transition to a spontaneously oscillating state in spin models.
A nonequilibrium generalization of the Landau free energy is obtained from the joint distribution of the magnetization and its smoothed stochastic time derivative. The order parameter of the transition is a Hamiltonian, whose nonzero value signals the onset of oscillations. The Hamiltonian and the nonequilibrium Landau free energy are determined explicitly from the stochastic spin dynamics.
The oscillating phase is also characterized by a non-trivial overlap distribution reminiscent of a continuous replica symmetry breaking, in spite of the absence of disorder.
An illustration is given on an explicit kinetic mean-field spin model.
\end{abstract}

\maketitle

%%%%%%%%%%%%%

% Introduction
The emergence of spontaneous oscillations at a collective scale in large assemblies of interacting units is one of the most striking features of nonequilibrium systems.
Beyond the now well-understood synchronization of coupled oscillators \cite{acebron_kuramoto_2005,risler_universal_2004},
spontaneous oscillations also appear in diverse systems of interacting units where individual units do not oscillate in the absence of interactions, making the onset of oscillations a genuinely collective phenomenon.
Such oscillations have been reported for instance in biochemical clocks \cite{Cao_free_energy2015,nguyen_phase_2018,Aufinger_complex2022},
%and chemical oscillators \cite{nicolis_dissipative_1986}, 
populations of biological cells \cite{Kamino_fold2017,Wang_emergence2019}, assemblies of active particles with non-reciprocal interactions \cite{saha_scalar_2020,you_nonreciprocity_2020}, nonequilibrium spin systems \cite{collet_rhythmic_2016,de_martino_oscillations_2019,daipra_oscillatory_2020} as well as population dynamics \cite{andrae_entropy_2010,Duan_Hopf2019} and socio-economic models \cite{Gualdi2015,yi_symmetry_2015}.

In the thermodynamic limit, the onset of spontaneous oscillations is described by a deterministic Hopf bifurcation \cite{crawford_introduction_1991}.
Yet, oscillations often occur in mesoscopic systems like biochemical clocks for which fluctuations play an important role \cite{Fei_design2018}, leading to a stochastic Hopf bifurcation \cite{Sagues2007,Xu_Langevin2020} and to a finite coherence time of oscillations \cite{gaspard_correlation_2002,barato_cost_2016,barato_coherence_2017,oberreiter_universal_2022,remlein_coherence_2022}.
%
%\cite{arnold_toward_1994} in the case of the noisy Duffing-van der Pol oscillator, \cite{xiao_effects_2007} in the case of chemical reaction dynamics
%
To provide a consistent theoretical ground, the emergence of spontaneous oscillations in large assemblies of interacting units has been characterized as a nonequilibrium thermodynamic phase transition, by identifying the entropy production as a generalized thermodynamic potential whose derivative is discontinuous at the transition
\cite{crochik_entropy_2005,xiao_entropy_2008,xiao_stochastic_2009,barato_entropy_2012,tome_entropy_2021,nguyen_phase_2018,noa_entropy_2019,martynec_entropy_2020,seara_irreversibility_2021}.
Similar results have also been obtained for the entropy production in population dynamics \cite{andrae_entropy_2010}, and for a nonequilibrium free energy in the context of Turing pattern formation \cite{falasco_information_2018}.
However, beyond singularities of thermodynamic potentials, the equilibrium theory of phase transitions and critical phenomena is based on the key concepts of spontaneous symmetry breaking and of associated order parameter \cite{LeBellac}.
Once the latter is identified, the generic Landau free-energy can be determined unambiguously to characterize the phase transition at mean-field level.
Recent nonequilibrium generalizations of Landau's theory include the description of relaxation effects \cite{Meibohm_2022,Holtzman_2022}, or multiple heat baths and oscillations driven by an oscillatory field \cite{Aron_2020}.

In this Letter, we go beyond the thermodynamic approach to phase transitions with spontaneously emerging oscillations, and show how to build a nonequilibrium generalization of the Landau free energy in a class of driven kinetic mean-field spin models, based on the spontaneous breaking of spin-reversal symmetry and time-translation invariance.
The generalized Landau free energy is obtained from the joint distribution of the magnetization and its smoothed stochastic time derivative, at odds with previous generalizations based on magnetization only \cite{Meibohm_2022,Holtzman_2022,Aron_2020}.
Close to the phase transition to an oscillating phase, the nonequilibrium Landau free energy can be expressed in terms of a single order parameter, which is an effective Hamiltonian describing the oscillating dynamics of the magnetization. %and its time derivative.
In addition, we show by evaluating the overlap distribution of spin configurations that the oscillating phase is also characterized by an analogue of the continuous replica symmetry breaking phenomenon observed in disordered systems \cite{mezard_spin_1987}.

%%%%%%%%%%%%%%%
%XXXXXXXXXXXXXXXXXXXX

%\paragraph{Spin models and stochastic derivative of the magnetization.}
We consider a generic class of nonequilibrium mean-field spin models with $N$ spins $s_i\pm1$ (and possibly auxiliary variables), and define the magnetization $m=N^{-1} \sum_{i=1}^N s_i$.
We explore far-from-equilibrium regimes where for large $N$ the magnetization $m(t)$ may exhibit oscillations, leading to a limit cycle
\cite{collet_macroscopic_2014,collet_rhythmic_2016,collet_effects_2019,de_martino_oscillations_2019,martino_feedback2019,daipra_oscillatory_2020}. 
In dynamical systems theory, a limit cycle may be generically described in the plane of a variable and its time derivative.
We aim at building a generalized Landau theory describing finite size fluctuations around the average limit cycle. We thus need to characterize not only the fluctuations of magnetization, but also of its time derivative. Yet, directly considering the time derivative of $m(t)$ leads to diverging, white-noise type fluctuations that are not appropriate to build a Landau theory. 
We thus rather aim at defining an observable attached to each microscopic configuration that would play the role of an appropriately smoothed out derivative of the magnetization. 
We denote as $\mC$ the microscopic configuration of the system; $\mC$ may correspond to the spin configuration $\mC=(s_1,\dots,s_N)$ \cite{collet_macroscopic_2014,collet_rhythmic_2016}, or may include additional binary variables, $\mC=(s_1,\dots,s_N,h_1,\dots,h_M)$, see below.
%(e.g., $M=1$ for feedback control \cite{de_martino_oscillations_2019} or $M=N$ in the explicit spin model below).
For a Markov jump dynamics with transition rate $W(\mC'|\mC)$ from configuration $\mC$ to configuration $\mC'$,
a stochastic derivative $\dot{m}(\mC)$ of the magnetization $m(\mC)$ can be defined as (see Supplemental Material \cite{SM})
    \begin{equation} \label{eq:def:mdot}
        \dot{m}(\mC)=\sum_{\mC'\neq \mC}\left(m\left(\mC'\right)-m\left(\mC\right)\right)W(\mC'|\mC).
    \end{equation}
This definition is such that $d\langle m \rangle/dt=\langle \dot{m} \rangle$,
where the average $\langle \dots \rangle$ is defined as $\langle x \rangle = \sum_{\mC} x(\mC)P(\mC)$.
%In the limit $N\to \infty$, the stochastic observable $\dot{m}$ converges to the deterministic derivative $dm/dt$.
The definition Eq.~\eqref{eq:def:mdot} of the derivative $\dot{m}$ is valid for any system size $N$ and leads to fluctuations on a scale comparable to that of $m$.

To break detailed balance and possibly allow for oscillations,
the configuration $\mC$ is split into two groups of binary variables denoted as $s_i^k$ ($k=a,b$) having different single-spin-flip dynamics (see \cite{SM} for details).
These may correspond to two groups of spins in contact with different heat baths \cite{lecomte_energy_2005,collet_macroscopic_2014,collet_rhythmic_2016},
or to the spin and field variables as in the explicit model described below.
To detect temporal oscillations, we use as global observables the magnetization $m$ and its stochastic time derivative $\dot{m}$ defined in Eq.~(\ref{eq:def:mdot}).
We consider the joint distribution $P_N(m, \dot{m})=\sum_{\mC\in\mathcal{S}(m, \dot{m})} P(\mC),$ where $\mathcal{S}(m, \dot{m})$ corresponds to the set of configurations $\mC$ with $m(\mC)=m$ and $\dot{m}(\mC)=\dot{m}$.
The coarse-grained transition rate corresponding to flipping any spin $s_i^k=\pm 1$ in group $k=a,b$,
starting from a configuration $\mC \in \mathcal{S}(m,\dot{m})$, is denoted as $N W_{k}^{\pm}(m,\dot{m})$. % in microscopic time units.
%$W_{k}^{\pm}(m,\dot{m})$ is thus the transition rate measured in macroscopic time units, after a rescaling of time $t \to t/N$.
A global spin-reversal symmetry is assumed, yielding $W_{k}^{\pm}(-m,-\dot{m})=W_{k}^{\mp}(m,\dot{m})$.
Variations of $m$ and $\dot{m}$ when flipping a spin $s_i^k=\pm 1$ ($k=a,b$) scale as $1/N$:
$(\Delta m, \Delta \dot{m})=\pm\mathbf{d}_{k}/N$.
The coarse-grained master equation governing the evolution of $P_N(m, \dot{m})$ reads
\be \begin{aligned}\label{eq:eq:p}
\partial_t& P_N(m, \dot{m})=N\sum_{k,\sigma} \bigg[ 
-W_k^{\sigma}(m, \dot{m})P_N(m, \dot{m})\\
&+W_k^{\sigma}\left((m, \dot{m})-\frac{\sigma\textbf{d}_{k}}{N}\right)P_N\left((m, \dot{m})-\frac{\sigma\textbf{d}_{k}}{N}\right)\bigg].
\end{aligned} \ee
From the theory of Markov jump processes with vanishing jump size \cite{knessl_1985}, the stationary joint distribution $P(m, \dot{m})$ takes for large $N$ a large deviation form \cite{touchette_2009}
    \begin{equation} \label{eq:large:dev:dist}
        P_N(m, \dot{m}) \sim \exp \left[-N\phi(m, \dot{m})\right],
    \end{equation}
which can be interpreted as a WKB approximation of the solution of the master equation (\ref{eq:eq:p}) \cite{knessl_1985}.
%The rate function $\phi(m,\dot{m})$ \cite{touchette_2009} may be considered as a nonequilibrium generalization of the Landau free energy, but it is now a function of the two variables $m$ and $\dot{m}$, characterizing respectively the spontaneous breaking of spin-reversal symmetry and of time-translation invariance.
Using the large deviation form (\ref{eq:large:dev:dist}) in Eq.~(\ref{eq:eq:p}) and taking the limit $N\to\infty$, one ends up with the following equation for the steady-state rate function $\phi(m,\dot{m})$ ,
\be \label{eq:HJ}
\sum_{k,\sigma} W_{k}^{\sigma}(m,\dot{m}) \left[ e^{\sigma\mathbf{d}_{k}\cdot\nabla\phi(m,\dot{m})} - 1 \right] = 0\,,
\ee
with $\nabla\phi=(\partial_m \phi, \partial_{\dot m} \phi)$.
%In a Landau theory spirit, 
We are interested in an expansion of $\phi(m,\dot{m})$ close to its minimum (or minima), and thus assume $\nabla\phi$ to be small.
At order $|\nabla\phi|^2$, Eq.~(\ref{eq:HJ}) reads 
\be \begin{aligned} \label{eq:phi:quadrat}
&\dot{m}\partial_m\phi + Y \partial_{\dot{m}}\phi
+ D_{11} (\partial_m\phi)^2 + D_{22} ( \partial_{\dot{m}}\phi)^2\\
&\qquad \qquad \qquad \qquad \qquad + D_{12} (\partial_m\phi)(\partial_{\dot{m}}\phi) = 0,
\end{aligned} \ee
where $Y$ and $\mathbf{D}=\{D_{ij}\}$ are defined as, using Eq.~(\ref{eq:def:mdot}),
\be \begin{aligned}
\big(\dot{m}, Y(m,\dot{m})\big) &= \sum_{k,\sigma} \sigma \mathbf{d}_{k} W_k^{\sigma}(m, \dot{m}),\\
\mathbf{D}(m,\dot{m}) &= \frac{1}{2}\sum_{k,\sigma} W_k^{\sigma}(m,\dot{m}) \, \mathbf{d}_{k}\!\cdot\! \mathbf{d}^T_{k}.
\end{aligned} \ee
At the transition to spontaneous oscillations, $\phi(m, \dot{m})$ should change from a paraboloid-like shape to a `Mexican-hat' shape.
To identify the parameter controlling the transition, we start with a quadratic approximation of $\phi(m,\dot{m})$ for small $m$ and $\dot{m}$, and look for a change of curvature.
At quadratic order in $m$ and $\dot{m}$, Eq.~(\ref{eq:HJ}) takes the same form as Eq.~(\ref{eq:phi:quadrat}), but with constant coefficients $D_{ij} \ge 0$ and a linear function
$Y(m,\dot{m})=-v_0 m + u_0 \dot{m}$, assuming $v_0>0$ ($Y(0,0)=0$ because $Y(-m,-\dot{m})=Y(m,\dot{m})$).
Assuming $\phi(m,\dot{m})=\frac{\gamma_1}{2}m^2+\frac{\gamma_2}{2}\dot{m}^2+\gamma_3 m\dot{m}$ with small $\gamma_i$'s close to the transition,
one finds $\gamma_3 \sim \gamma_1^2 \ll \gamma_1$ and $u_0 \gamma_2 = -(D_{11} \gamma_1^2 /v_0 + D_{22} \gamma_2^2) <0$.
The sign of $\gamma_2 = \partial^2\phi/\partial \dot{m}^2(0,0)$ is thus the opposite of the sign of $u_0$.
Hence $u_0$ is the control parameter of the phase transition: $u_0=0$ corresponds to the critical point, and time-translation invariance is broken for $u_0>0$, when $\dot{m}=0$ is no longer stable.

For $u_0>0$, the quadratic approximation is not enough to describe the minima of $\phi(m, \dot{m})$, and higher order terms are required.
One could expand $\phi(m, \dot{m})$ as a power series in $m$ and $\dot{m}$, but this would not work for nonanalytic $\phi$ [see, e.g., Eq.~(\ref{eq:f:H:LC:nonelliptic})].
Instead, we use the Hamiltonian structure close to the critical point.
We no longer assume $Y(m,\dot{m})$ to be linear, and split $Y(m,\dot{m})$ into the $\dot{m}$-independent part $Y(m, 0)\equiv -V'(m)$ and a $\dot{m}$-dependent part $Y(m, \dot{m})-Y(m, 0)\equiv \dot{m}g(m, \dot{m})$.
We define the control parameter $u_0$ as $u_0=\partial Y/\partial\dot{m}(0,0)$.
We take $u_0\propto\ve$ with $\ve$ a small parameter. To perform a consistent small-$\ve$ expansion of Eq.~(\ref{eq:HJ}), we assume 
$\nabla \phi = O(\ve)$, since quadratic terms in $\nabla \phi$ have to balance the contribution in $\ve \partial_{\dot{m}}\phi$ coming from the term $Y \partial_{\dot{m}}\phi$.
Truncating Eq.~(\ref{eq:HJ}) at order $\ve^2$, one recovers Eq.~(\ref{eq:phi:quadrat}), where the full ($m,\dot{m}$)-dependence of the coefficients is kept.
At order $\epsilon$, Eq.~(\ref{eq:phi:quadrat}) reduces to 
\be \label{eq:phi:linear}
\dot{m}\partial_m\phi -V'(m) \partial_{\dot{m}}\phi =0.
\ee
The general solution of Eq.~(\ref{eq:phi:linear}) reads
\be \label{eq:phi:fH}
\phi(m,\dot{m}) = f\big(H(m,\dot{m})\big) + f_0
\ee
with
\be \label{eq:def:H}
H(m, \dot{m}) = \frac{\dot{m}^2}{2} + V(m),
\ee
and where $f$ is at this stage an arbitrary function, satisfying for convenience $f(0)=0$,
and the constant $f_0$ ensures that the minimal value of $\phi(m,\dot{m})$ is zero.
The minimum value of $V(m)$ is set to $V=0$, so that $H\ge 0$.
$H(m, \dot{m})$ is a Hamiltonian describing the $(m,\dot{m})$ dynamics at order $\ve$ as
$\frac{dm}{dt}=\frac{\partial H}{\partial \dot{m}}$, $\frac{d\dot{m}}{dt}=-\frac{\partial H}{\partial m}$,
and the corresponding trajectories are iso-$\phi$ lines.
Contributions of order $\ve^2$ to Eq.~(\ref{eq:phi:quadrat}) yield a condition determining the derivative $f'(H)$ \cite{SM}, 
\be \label{eq:fprime}
f'(H) = -\frac{\int_{m_1}^{m_2}dm\, \dot{m}(m, H)\, g\big(m,\dot{m}(m, H)\big)}
  {\int_{m_1}^{m_2}\frac{dm}{\dot{m}(m, H)} \,\nabla^TH\cdot \mathbf{D} \cdot \nabla H}
\ee
%with a matrix $\mathbf{D}=\{D_{ij}\}$ given by
%\be
%\mathbf{D}=\frac{1}{2}\sum_k W_k(m, \dot{m}) \, \mathbf{a}_{k}\!\cdot\! \mathbf{a}^T_{k},
%\ee
where $m_1$ and $m_2$ are such that
$V(m_1)=V(m_2)=H$ and $V(m)\le H$ for $m_1\le m\le m_2$; $\dot{m}(m, H)$ is determined from Eq.~(\ref{eq:def:H}).
%$\dot{m}(m, H)=\sqrt{2(H-V(m))}$;
%and $\nabla^TH\cdot D\cdot \nabla H= D_{11}V'(m)^2+2D_{12}V'(m)\dot{m}(m, H)+D_{22}\dot{m}(m, H)^2$.
Note that a related method has been used to determine nonequilibrium potentials in dissipative dynamical systems \cite{Graham_nonequilibrium1987, graham.tel.1984a, graham_1989}.

Eqs.~(\ref{eq:phi:fH}) and (\ref{eq:fprime}) provide a convenient description of a mean-field phase transition to a state with temporal oscillations.
The function $f(H)$ plays a role similar to the Landau free energy at equilibrium. Let us denote as $H^*$ the value of $H$ which minimizes $f(H)$.
The case $H^*=0$ corresponds to usual time-independent phases, either paramagnetic or ferromagnetic depending on whether $V(m)$ is minimum for $m=0$ or $m\ne 0$ respectively. 
The case $H^*>0$ instead corresponds to the onset of spontaneous oscillations, where $(m,\dot{m})$ follow a limit cycle in the deterministic limit
$N\to\infty$.
Hence $H^*$ may be considered as the formal order parameter of the transition to an oscillating state.
%(other order parameters may be more convenient in practice, see below).
Note that although the system exhibits macroscopic temporal oscillations, the probability distribution $P_N(m,\dot{m})$ is time-independent (in the long-time limit), because it describes an infinite ensemble of systems oscillating at the same frequency, but with uniformly distributed phases.

%A key point is that the effective potential $V(m)$ intervenes in Eq.~(\ref{eq:phi:fH}) in both the definition of the Hamiltonian $H$ and the function $f(H)$, through its derivative $f'(H)$ given by Eq.~(\ref{eq:fprime}). Hence the functional forms of $H(m,\dot{m})$ and of $f(H)$ cannot be decoupled.
In the simple yet generic case where $V(m)=\frac{1}{2}v_0 m^2$ and $g(m,\dot{m})=\alpha_0\ve-\alpha_{1}m^2-\alpha_{2}m\dot{m}-\alpha_{3}\dot{m}^2$, $f(H)$ takes for small $H$ the generic form
\be \label{eq:f:H:LC:elliptic}
f(H)=-\ve aH+bH^2,
\ee
where %$\ve$ is the control parameter of the transition, and 
$a$ and $b$ can be expressed in terms of the parameters $\alpha_i$ \cite{SM}. The case $\ve<0$ corresponds to a time-independent phase ($H^*=0$),
while $\ve>0$ corresponds to an oscillating phase, with $H^*=\ve a/2b>0$.
One thus finds a continuous phase transition to temporal oscillations, with
an elliptic limit cycle whose size scales as $\ve^{1/2}$, i.e., $m\sim\dot{m}\sim\ve^{1/2}$, or more precisely $\la m^2\ra \sim \la \dot{m}^2\ra \sim \ve$.
The two observables $\la m^2\ra$ and $\la \dot{m}^2\ra$ constitute the practically measurable order parameters, respectively characterizing
the paramagnetic-ferromagnetic phase transition and the onset of spontaneous oscillations.
%of phase transitions occurring in spin models, as they can be easily measured in a numerical simulation:
%$\la m^2\ra$ characterizes the paramagnetic-ferromagnetic phase transition, while $\la \dot{m}^2\ra$ characterizes the emergence of temporal oscillations in steady state.
From the expression \eqref{eq:def:H} of the Hamiltonian $H$,
the oscillation period $\tau$ is given in the case $V(m)=\frac{1}{2}v_0 m^2$ by $\tau=2\pi/\sqrt{v_0}$, and is thus independent of $\ve$.
Yet, the scaling with $\ve$ of the different observables may differ from the results given above. Close to a tricritical point where the paramagnetic, ferromagnetic and oscillating phases meet, one rather finds
$V(m)=\frac{1}{4}v_1 m^4$ (see explicit example below).
In this case, $f(H)$ takes the nonanalytic form
\be \label{eq:f:H:LC:nonelliptic}f(H)=-\ve a H+cH^{3/2}\ee from Eq.~(\ref{eq:fprime}) \cite{SM}, and the scaling of $H^*$ is now $H^* \sim \ve^2$ instead of $H^*\sim \ve$. As $V(m)$ is proportional to $m^4$, $m$ and $\dot{m}$ have different scalings with $\ve$: $m\sim \ve^{1/2}$, while $\dot{m} \sim \ve$. The limit cycle is no longer elliptic but it flattens. This actually corresponds to a period that diverges as
$\tau\sim \ve^{-1/2}$. 

%The advantage of the description in terms of the rate function
%$\phi(m,\dot{m})$ is twofold. First, it allows for a characterization of macroscopic fluctuations of $m$ and $\dot{m}$ around the deterministic limit cycle for large but finite $N$.
The small fluctuations of $m$ and $\dot{m}$ around their zero average value in the paramagnetic phase $\ve<0$ can be characterized by generalized susceptibilities $\chi_{m} = N\langle m^2 \rangle$ and $\chi_{\dot{m}} = N\langle \dot{m}^2\rangle$, taking into account that $\langle m^2 \rangle \sim \langle \dot{m}^2\rangle \sim N^{-1}$ in the paramagnetic phase.
%
%Note that
%at odds with $\chi_{m}$ which at least at equilibrium is related to the linear response to an external field,
%$\chi_{\dot{m}}$ cannot be related in a simple way to a response to a field, but is only defined in terms of fluctuations.
%$\chi_{\dot{m}} \sim \ve^{-1}$ and $\chi_{m} \sim \ve^{-1}$.
When approaching the phase transition to a limit cycle ($\ve\to 0^{-}$), both generalized susceptibilities $\chi_{\dot{m}}$ and $\chi_{m}$ diverge as $\vert\ve\vert^{-1}$.
At the critical point ($\ve=0$), one finds a different scaling of fluctuations with $N$: $\langle \dot{m}^2 \rangle \sim \langle m^2 \rangle \sim  N^{-1/2}$.
As for the finite-size fluctuations of $H$, we obtain that in the paramagnetic phase, $\text{var}(H)\sim N^{-2}$ whereas in the oscillating phase $\text{var}(H)\sim N^{-1}$.

The rate function is a key tool to determine which solution is the macroscopically observed one when two or more solutions are present in the deterministic description.
This is the case, e.g., when $f(H)=aH-bH^2+cH^3$, % (which is possible for $V(m)=\frac{1}{2}v_0m^2$)
with $a$, $b$, $c>0$. Both $H^*=0$ and $H^*=\frac{b+\sqrt{b^2-3ac}}{3c}>0$ are local minima of $f(H)$, corresponding to two solutions of the deterministic equations. 
The macroscopically observed solution is the one with the lowest $f(H)$.
Varying parameters, one thus observes a discontinuous transition from a paramagnetic phase ($H^*=0$) to a limit cycle phase ($H^*>0$).
An explicit example is given below.

\begin{figure}[t]
    \centering
    \includegraphics[width=\columnwidth]{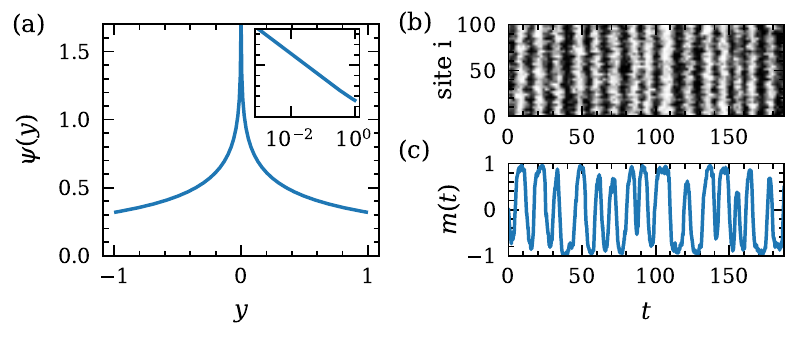}
    \caption{(a) The scaling function $\psi(y)$ of the overlap distribution in the oscillating phase. The inset corresponds to a logarithmic $x$-axis. (b) Evolution of the $N$ spins with time
    ($s_i=+1$ in white, $s_i=-1$ in black) obtained using Monte Carlo simulations of the specific model described below, for $(T_c-T)/T_c=0.3$, $\mu=2$, and $N=100$. (c) The corresponding $m(t)=N^{-1}\sum s_i$
    vs.~time $t$.}
    \label{fig:mt_psiy}
\end{figure}

%\paragraph{Overlap between spin configurations.}
A fine characterization of the phase transition to an oscillating state is obtained by considering the statistics of the overlap $q_{ab}=N^{-1}\sum_{i=1}^N s_i^{a} s_i^{b}$ between two spin configurations $\{s_i^{a}\}$ and $\{s_i^{b}\}$.
%Overlaps have been introduced in spin-glass models to deal with the absence of a visible order in the spin-glass phase \cite{mezard_spin_1987}.
Identical (opposite) configurations have an overlap $q_{ab}=1$ ($q_{ab}=-1$),
while $q_{ab}=0$ for uncorrelated configurations.
%For instance, in the Ising model in the ferromagnetic phase, only configurations with a magnetization $m=\pm m_0$ are explored for $N\to\infty$ (where $m_0$ is the spontaneous magnetization), and two randomly chosen configurations typically have a non-zero overlap $q=\pm m_0^2$.
The overlap distribution $P(q)$ can be evaluated for $N\to \infty$ \cite{SM}.
%based on the spin-configuration distribution $P(\{s_i\})$
%
%In  the paramagnetic-oscillating phase transition with an elliptic limit cycle
For $V(m)=\frac{1}{2}v_0 m^2$, we obtain %for $\epsilon <0$ (paramagnetic phase) $P(q)=\delta(q)$, and 
for $\epsilon>0$ (oscillating phase) the scaling form $P(q)=q_{\ve}^{-1} \psi(q/q_\ve)$, with $q_{\ve}=\ve a/bv_0$ [$a$ and $b$ are introduced in Eq.~\eqref{eq:f:H:LC:elliptic}]; the scaling function $\psi(y)$ is plotted in Fig.~\ref{fig:mt_psiy}(a) (see \cite{SM} for its explicit expression).
%    \begin{equation} \label{eq:psi:scalfn}
%      \psi(y) = \theta(1-|y|)\frac{2}{\pi^2} \int_{|y|}^{1} \frac{\mathrm{d}x}{\sqrt{(1-x^2)(x^2-y^2)}} 
%    \end{equation}
%where $\theta$ is the Heaviside function. The scaling function $\psi(y)$ is plotted in Fig.~\ref{fig:mt_psiy}.
$P(q)$ has a logarithmic divergence in $q=0$, and %reaches a constant value $bv_0/\pi\ve a$ when $|q|\to q_{\ve}$. The probability density of the overlap 
has a continuous support, a property usually considered as a hallmark of continuous replica symmetry breaking in disordered systems \cite{mezard_spin_1987}.
%Here, no disorder is present in the system, and a replica symmetry breaking is thus not expected.
As in the latter, the presence of a non-trivial overlap distribution can be traced back to an average over many pure states
\cite{mezard_spin_1987,guislain23}.

\begin{figure}[t]
    \centering
    \includegraphics{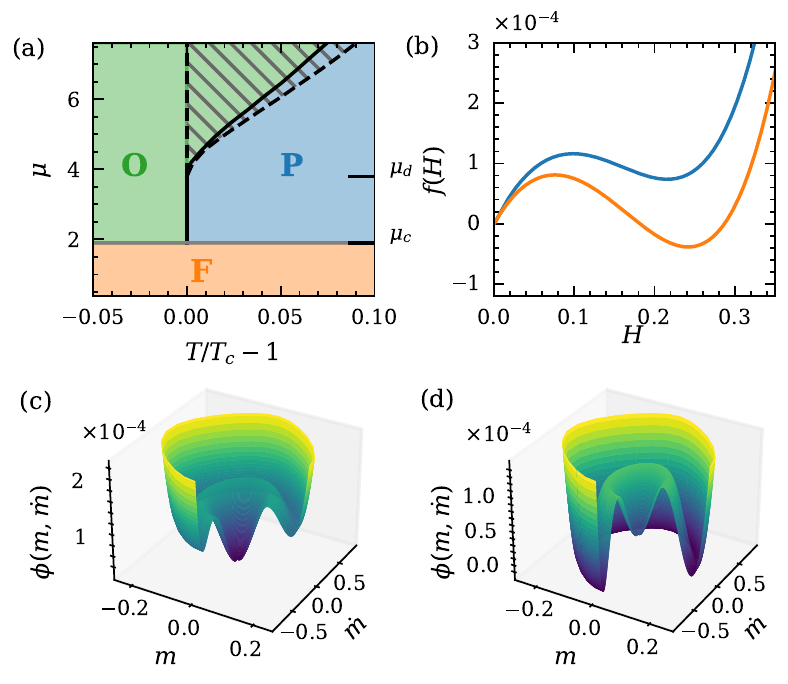}
    \caption{(a) Phase diagram of the spin model in the $(-\ve,\mu)$-plane, with $\ve=(T_c-T)/T_c$, displaying the paramagnetic (P), ferromagnetic (F) and oscillating (O) phases
    ($J_1=1.4$, $J_2=-0.5$). %$\mu_c=1.9$, $\mu_d=3.8$, $T_c=0.45$).
    P and O phases coexist in the hatched area.
    (b) $f(H)$ for $\mu=5.7$ and  $\ve=-3.5\times10^{-2}$ (top curve), $\ve=-3.0\times10^{-2}$ (bottom curve). 
    (c), (d) Corresponding rate function $\phi(m, \dot{m})$.}%(e) and (f) Rate function $\phi(0, \dot{m})$ for the same parameters as (c) and (d) respectively. }
    \label{fig:phase:diag:f:phi}
\end{figure}

As an explicit model, we introduce a generalization of the kinetic mean-field Ising model with ferromagnetic interactions
(see also related models with two spin populations \cite{collet_macroscopic_2014,collet_rhythmic_2016}
or with feedback control \cite{de_martino_oscillations_2019}). 
The model involves $2N$ microscopic variables: $N$ spins $s_i=\pm1$ and $N$ fields $h_i=\pm 1$. 
We define the magnetization $m = N^{-1} \sum_{i=1}^N s_i$ and
the average field $h=N^{-1} \sum_{i=1}^N h_i$.
The stochastic dynamics consists in randomly flipping a single spin $s_i$ or a single field $h_i$. The flipping rates $W_s$ and $W_h$ depend only on $m$ and $h$, $W_{s,h} = [1+\exp(\beta \Delta E_{s,h})]^{-1}$, with $\beta=T^{-1}$ the inverse temperature and $\Delta E_{s,h}$ the variation of $E_{s,h}$ when flipping a spin $s_i$ or a field $h_i$, where $E_s=-N(\frac{J_1}{2} m^2+\frac{J_2}{2} h^2+mh)$ and $E_h=E_s+\mu Nhm$.
Detailed balance is broken as soon as $\mu \ne 0$.
%For $N\to\infty$, $m$ and $h$ obey deterministic equations,
%\begin{align}
%  \label{eq:dyn:m}
%  \dot{m} &= -m+\tanh[\beta(J_1 m+h)],\\
%  \label{eq:dyn:h}
%  \dot{h} &= -h+\tanh[\beta(J_2 h+(1-\mu)m)].
%\end{align}
The fluctuating derivative $\dot{m}$ determined from Eq.~(\ref{eq:def:mdot}) reads
$\dot{m} = -m+\tanh[\beta(J_1 m+h)]$. 

Depending on $(T,\mu)$ values, the model exhibits a paramagnetic (high $T$), ferromagnetic (low $T$, low $\mu$) or oscillating (low $T$, high $\mu$) behavior. We restrict the study to $J_1>-J_2$. An example of a phase diagram is shown in %A typical phase diagram for $1<J_1<1-J_2$ and $-1<J_2<0$ is shown in 
Fig.~\ref{fig:phase:diag:f:phi}(a) for $J_1=1.4$ and $J_2=-0.5$. The boundary of the ferromagnetic phase is obtained from the deterministic equations \cite{SM}. Other lines are obtained using the perturbative framework introduced in Eqs.~\eqref{eq:fprime} and \eqref{eq:phi:fH} \cite{SM}.
%(see \cite{SM} for the analytic expressions of the different functions used to compute the rate function). 
%In this example, $Y(m, \dot{m})$ is analytic and can be expanded in power series of $m$ and $\dot{m}$, allowing us to use previous analytical results. 
%The function $f(H)$ can be evaluated numerically from Eq.~\eqref{eq:phi:fH}, and its global minimum gives the most stable phase. 
%
The three phases meet at a tricritical point $(T_c,\mu_c)$, with
$T_c=\frac{J_1+J_2}{2}$ and $\mu_c=1+\frac{(J_1-J_2)^2}{4}$.
For $\mu_c<\mu<\mu_d$, where $\mu_d=1-\frac{J_1}{J_2}$, a continuous transition from paramagnetic to oscillating states (with an elliptic limit cycle) is observed. 
An example of the oscillations of the $N$ spins $s_i$ with time and $m(t)=N^{-1}\sum_i s_i$, obtained from Monte-Carlo simulations in the oscillating phase is plotted in Fig.~\ref{fig:mt_psiy}(b) and (c).
The rate function obtained numerically from Eq.~\eqref{eq:fprime} is well described by Eq.~\eqref{eq:f:H:LC:elliptic}, with a reduced control parameter $\ve =(T_c-T)/T_c$. 
Close to the tricritical point ($\mu \gtrsim \mu_c$), an elongated limit cycle is observed, with $m \sim \ve^{1/2}$ and $\dot{m} \sim \ve$.
Here, the rate function is instead well described by the nonanalytic form of $f(H)$ obtained in Eq.~\eqref{eq:f:H:LC:nonelliptic} (the value of $c$ is given in \cite{SM}).
For $\mu>\mu_d$, a discontinuous transition from paramagnetic to oscillating states is observed. In the hatched area of Fig.~\ref{fig:phase:diag:f:phi}(a), both the paramagnetic ($H^*=0$) and limit cycle ($H^*>0$) states are local minima of $f(H)$. The most stable solution at large but finite $N$ is then determined as the global minimum of $f(H)$, see Fig.~\ref{fig:phase:diag:f:phi}(b). It discontinuously changes from $H^*=0$ (paramagnetic state) to $H^*>0$ (oscillating state) when crossing the full line inside the hatched area of Fig.~\ref{fig:phase:diag:f:phi}(a). The rate function $\phi(m,\dot{m})$ is plotted in Fig.~\ref{fig:phase:diag:f:phi}(c) and (d) %Fig.~\ref{fig:phase:diag:f:phi}(d) 
for the paramagnetic and oscillating states respectively. The metastable (oscillating or paramagnetic) states are also visible.
Note that the validity of the perturbative framework is limited to small $(T_c-T)/T_c$ and to either $\mu_c<\mu<\mu_d$ or small $(\mu-\mu_d)/\mu_d>0$.
A detailed study of this model, including a description of the transition between ferromagnetic and limit cycle states, will be reported elsewhere \cite{guislain23}.

To sum up, we have shown how the Landau theory of phase transitions can be extended to describe phase transitions to an oscillating phase in nonequilibrium spin models.
While previous nonequilibrium generalizations of the Landau free energy were only based on magnetization and did not address spontaneous oscillations \cite{Meibohm_2022,Holtzman_2022,Aron_2020}, we defined a generalized Landau free energy as the rate function $\phi(m,\dot{m})$ associated with the joint distribution of the magnetization $m$ and its smoothed stochastic derivative $\dot{m}$ defined in Eq.~(\ref{eq:def:mdot}).
The order parameter of the Landau theory is an effective Hamiltonian $H$, whose nonzero value indicates the presence of oscillations. The expression of $H(m,\dot{m})$ and of the nonequilibrium Landau free energy $f(H)$ can be determined explicitly from the stochastic spin dynamics. The expansion of $f(H)$ is singular close to a tricritical point where paramagnetic, ferromagnetic and oscillating phases meet. Beyond spontaneous breaking of time translation invariance, the oscillating phase is characterized by an overlap distribution reminiscent of continuous replica symmetry breaking, although no disorder is present. Consistently with previous works \cite{crochik_entropy_2005,xiao_entropy_2008,xiao_stochastic_2009,barato_entropy_2012,tome_entropy_2021,nguyen_phase_2018,noa_entropy_2019,martynec_entropy_2020,seara_irreversibility_2021}, we also recover that the entropy production density becomes non-zero in the oscillating phase \cite{SM}.
Future work will notably aim at characterizing the transition to oscillating states in finite-dimensional systems using renormalization group methods.

%\acknowledgments


\newpage 
%\title{test}
%\maketitle

\begin{widetext} 
  \begin{center}\textbf{\large Supplementary Information: Nonequilibrium phase transition to temporal oscillations in mean-field spin models}\end{center}
\end{widetext}

\setcounter{equation}{0}
\setcounter{figure}{0}

\section{Generic spin model}

\subsection{Definition of the stochastic derivative of the magnetization}
We aim at defining a random variable that would play the role of the derivative of the magnetization. 
We denote as $\mC$ the microscopic configuration of the system. We assume a Markov jump dynamics with transition rate $W(\mC'|\mC)$ from configuration $\mC$ to configuration $\mC'$.
We introduce a stochastic derivative $\dot{m}(\mC)$ of the magnetization $m(\mC)$ such that in average, $d\la m\ra /dt=\la \dot{m}\ra$ where the average $\langle \dots \rangle$ is defined $\la x\ra =\sum_C x(\mC)P(\mC)$.
Using the master equation
\be \label{eq:SM:me}\frac{d P(\mC)}{dt}=\sum_{\mC'\neq \mC} \big[ W(\mC|\mC')P(\mC')-W(\mC'|  \mC)P(\mC) \big], \ee
we obtain after rearranging terms that 
\be \label{eq:def:mdot:av:SM}
\frac{d\langle m \rangle}{dt}=\sum_\mC  P(\mC)\sum_{\mC'\neq \mC} \left(m(\mC')-m(\mC)\right)W(\mC'| \mC).
\ee
From Eq.~(\ref{eq:def:mdot:av:SM}), we identify the stochastic derivative of the magnetization 
\begin{equation} \label{eq:def:mdot:SM}
    \dot{m}(\mC)=\sum_{\mC'\neq \mC}\left(m\left(\mC'\right)-m\left(\mC\right)\right)W(\mC'|\mC),
\end{equation}
in such a way that
\be \label{eq:SM:dt:av:m:is:mdot}
\frac{d\langle m \rangle}{dt}=\sum_\mC  P(\mC)\, \dot{m}(\mC).
\ee
The definition Eq.~(\ref{eq:def:mdot:SM}) of $\dot{m}$ provides a smoothed expression of the time derivative of $m$, in the sense that it is already averaged over possible arrival configurations $\mC'$. Fluctuations of $\dot{m}$ are thus on the same scale as that of $m$, which is appropriate to define a joint probability distribution of $m$ and $\dot{m}$ and its associated large deviation function.

\subsection{Transition rates in  the spin model}
\label{sec:trans:rate}

The microscopic configuration $\mC$ is split into two groups of binary variables having different dynamics, $\mC=(\mC_a,\mC_b)$ with $N_a$ and $N_b$ variables respectively.
These may correspond for instance to two groups of spins in contact with different heat baths,
$\mC_a=(s_1,\dots,s_{N_a})$ and $\mC_b=(s_{N_a+1},\dots,s_{N_a+N_b})$ with $N_a+N_b=N$
\cite{lecomte_energy_2005_SM,collet_macroscopic_2014_SM,collet_rhythmic_2016_SM},
or to the spin and auxiliary field variables as in the explicit model described in the main text, $\mC_a=(s_1,\dots,s_N)$ and $\mC_b=(h_1,\dots,h_M)$, in which case $N_a=N$ and $N_b=M$.

To keep notations generic, we write $\mC_a=(s_1^a,\dots,s_{N_a}^a)$ and $\mC_b=(s_1^b,\dots,s_{N_b}^b)$, and call $s_i^k$ a spin ($i=1,\dots,N_k$; $k=a,b$).
A single-spin-flip stochastic dynamics is assumed.
For a given $s_i^k=\pm 1$, the flipping rate $w^{\pm}_k(m_a,m_b)$ is independent of $i$ and depends only on the group magnetizations 
\be
m_{k'} = \frac{1}{N_{k'}} \sum_{j=1}^{N_{k'}} s_j^{k'}, \qquad k'=a,b.
\ee
This results from the mean-field assumption that the flipping rates are invariant under arbitrary spin permutations in each group $\mC_a$ and $\mC_b$.

To parametrize transition rates, we use the magnetization $m$ and its stochastic time derivative $\dot{m}$ defined in Eq.~(1) of the main text, instead of $(m_a,m_b)$, 
since the observables $(m,\dot{m})$ are better suited to detect temporal oscillations.
We thus reexpress $w^{\pm}_k(m_a,m_b)$ as $\tilde{w}^{\pm}_k(m,\dot{m})$.
%We consider the joint distribution $P_N(m, \dot{m})=\sum_{\mC\in\mathcal{S}(m, \dot{m})} P(\mC),$ where $\mathcal{S}(m, \dot{m})$ corresponds to the set of configurations $\mC$ with $m(\mC)=m$ and $\dot{m}(\mC)=\dot{m}$.
The coarse-grained transition rate corresponding to flipping any spin $s_i^k=\pm 1$ in a given group $k=a,b$,
starting from the configuration $\mC \in \mathcal{S}(m,\dot{m})$, is denoted as $N W_{k}^{\pm}(m,\dot{m})$ in microscopic time units,
with 
\be \label{eq:def:Wk}
W_{k}^{\pm}(m,\dot{m})=n_k^{\pm} \, \tilde{w}^{\pm}_k(m,\dot{m}),
\ee
where $n_k^{\pm}=\frac{1}{2}(1\pm m_k)$ is the fraction of spins $s_i^k=\pm 1$ in group $k=a,b$,
which can be reexpressed as a function of $(m,\dot{m})$.
$W_{k}^{\pm}(m,\dot{m})$ is thus the transition rate measured in macroscopic time units, after a rescaling of time $t \to t/N$.
%The variations of $m$ and $\dot{m}$ when flipping a spin $s_i^k=\pm 1$ ($k=a,b$) scale as $1/N$: $(\Delta m, \Delta \dot{m})=\mathbf{d}_{k,\pm}/N$.
%The coarse-grained master equation governing the evolution of $P_N(m, \dot{m})$ is then given by Eq.~(2) of the main text.
%
\subsection{Expression of the stochastic derivative of the magnetization in the spin model}
Starting from a configuration $C$ with $m(C)=m$ and $\dot{m}(C)=\dot{m}$, there are $N n_k^{\pm}$ possibilities to flip a spins $s_i^k=\pm$ in a given group $k=a, b$. By definition, we have $m(\mC')-m(\mC)=\pm d_k^1/N$ with the notations $\mathbf{d}_k=(d_k^1,d_k^2)$ and $W(C'|C)=\tilde{w}_k^{\pm}(m, \dot{m})$, so that Eq.~(\ref{eq:def:mdot:SM}) becomes
\begin{equation}
\dot{m}=\sum_{k, \sigma}\sigma d_{k}^1 n_k^{\pm}\tilde{w}_k^{\pm}(m, \dot{m}),
\end{equation}
and from Eq.~(\ref{eq:def:Wk}) above, one can identify $\dot{m}$ with the term $\sum_{k, \sigma}\sigma d_{k}^{1}W_k^{\sigma}(m, \dot{m})$ as indicated in Eq.~(6) of the main text.

\subsection{Deterministic equations}
As done for $\langle m\rangle$ in Eq.~(\ref{eq:SM:dt:av:m:is:mdot}), one can obtain an equation on the time derivative of $\langle \dot{m}\rangle$ 
\be\label{eq:dmdot:dt:average:SM} \frac{d\langle \dot{m}\rangle}{dt}=\sum_{\mC}P(\mC)\sum_{\mC'\neq \mC} (\dot{m}(\mC')-\dot{m}(\mC))W(\mC'|\mC).\ee
Using the notations introduced in the main text, the right hand side of this equation becomes 
\begin{equation}
\left\langle \sum_{k, \sigma}\sigma d_{k}^2W_k^{\sigma}(m, \dot{m})\right\rangle=\left\langle Y(m, \dot{m})\right\rangle.
\end{equation}
Therefore, assuming that the law of large numbers applies in the limit $N\to \infty$, the deterministic evolution equations on $m$ and $\dot{m}$ is 
\be   \label{eq:sys:der:m:mdot} \frac{dm }{dt}= \dot{m}, \quad     \frac{d \dot{m}}{dt}=Y(m, \dot{m}).\ee

%\subsection{Deterministic equations}
%Similarly, one can obtain an equation on the time derivative of $\langle \dot{m}\rangle$, 
%\be \frac{d\langle \dot{m}\rangle}{dt}=\sum_{\mC}P(\mC)\sum_{\mC'\neq \mC} (\dot{m}(\mC')-\dot{m}(\mC))W(\mC'|\mC).\ee
%Regrouping by transitions of type $k$, one gets
%\be \frac{d\langle \dot{m}\rangle}{dt}=\sum_{\mC}P(\mC) \sum_k a_k^2 W_k(m(\mC), \dot{m}(\mC)).\ee
%Using Eq.~(6) of the main text, one gets the following system,
%\be   \label{eq:temporal:der:average:m:mdot} \frac{d\langle m \rangle}{dt}=\langle \dot{m}\rangle, \quad     \frac{d\langle \dot{m}\rangle}{dt}=\langle Y(m, \dot{m})\rangle.\ee
%Assuming that the law of large numbers applies in the limit $N\to \infty$, the deterministic system on $m$ and $\dot{m}$ is given by Eq.~(XX) of the main text.

\section{Large deviations}

\subsection{Determination of $f'(H)$}
To obtain the large deviation function $\phi(m, \dot{m})$ we adapt to stochastic models of interacting spins a method presented in \cite{Graham_nonequilibrium1987_SM} in the context of dissipative dynamical systems weakly perturbed by noise. 

%From the master equation and the form of the large deviation function, we obtain the following equation on $\phi(m, \dot{m})$
%\be \label{eq:eq:phi}\sum_{k} W_{k}(m,\dot{m}) \left[ e^{\mathbf{a}_{k}\cdot\nabla\phi(m,\dot{m})}
%  - 1 \right] = 0.\ee
%The leading order in $\ve$ of Eq.~\eqref{eq:eq:phi} gives
%\be \label{eq:eq:phi:order:1} \dot{m}\partial_m\phi-V'(m)\partial_{\dot{m}}\phi=0.\ee
%Introducing the Hamiltonian 
%$ H(m, \dot{m})=\frac{1}{2}\dot{m}^2+V(m)$ which is solution of  $\dot{m}\partial_m H-V'(m)\partial_{\dot{m}}H=0$, we obtain a family of solution of Eq.~\eqref{eq:eq:phi:order:1} as
%\be \phi(m, \dot{m})=f(H(m, \dot{m}))+f_0\ee
%where $f$ is an arbitrary function and $f_0$ a constant such that the minimal value of $\phi(m, \dot{m})$ is zero. 

From Eq.~(8) of the main text, the large deviation function $\phi(m, \dot{m})$ is expressed in terms of an unknown function $f(H)$.
To determine the function $f$, we consider the contribution at order $\varepsilon^2$ in the $\varepsilon$ expansion of Eq.~(5) of the main text, which reads
\be \label{eq:eq:phi:order:2_SM} \begin{aligned}
0&=\dot{m}\partial_m\phi_2 -V'(m)\partial_{\dot{m}}\phi_2\\
&+\dot{m}^2g(m, \dot{m})f'(H)+\left(\nabla^T H \cdot \mathbf{D} \cdot \nabla H\right)f'(H)^2
\end{aligned}\ee
with $\mathbf{D}$ given in Eq.~(6) of the main text, and
%$D=\frac{1}{2}\sum_k W_k(m, \dot{m})\mathbf{a}_k\cdot\mathbf{a}_k^T$; 
\be
\nabla^T H \cdot \mathbf{D} \cdot \nabla H = D_{11}V'(m)^2+2D_{12}V'(m)\dot{m}+D_{22}\dot{m}^2.
\ee
The first two terms of this equation depend on the contribution of order $\ve^2$ of $\phi$, which we note $\phi_2$. The last term depends on $f$, the leading order contribution to $\phi(m,\dot{m})$ in the $\varepsilon$ expansion.

We consider a closed trajectory of the Hamiltonian (constant $H$). We introduce $s$, a coordinate along this trajectory such that $\frac{dm}{ds}=\frac{\partial H}{\partial \dot{m}}$ and $\frac{d\dot{m}}{ds}=-\frac{\partial H}{\partial m }$.
The choice $H=V(m)+\frac{\dot{m}^2}{2}$ gives $\frac{dm}{ds}=\dot{m}$. The coordinate $s$ can thus be identified with time $t$. 

We define $m_1$ and $m_2$ such that $V(m_1)=V(m_2)=H$, and $s_0$ the coordinate such that $s=0$ and $s=s_0$ both correspond to the point $(m=m_1, \dot{m}=0)$. 
Integrating Eq.~\eqref{eq:eq:phi:order:2_SM} over $s$, the first term vanishes,
\be \int_{0}^{s_0}ds \left( \dot{m}\partial_m\phi_2 -V'(m)\partial_{\dot{m}}\phi_2\right)= \int_{0}^{s_0}ds\frac{d\phi_2}{ds}=0\ee
and using 
\be \int_{0}^{s_0}ds=2\int_{m_1}^{m_2}\frac{dm}{|\dot{m}(m, H)|} \ee
the second term of Eq.~\eqref{eq:eq:phi:order:2_SM} integrated and divided by $2f'(H)$, gives 
\be 0=\int_{m_1}^{m_2}dm\, \dot{m}g(m, \dot{m})+f'(H)\int_{m_1}^{m_2} dm\, \frac{\nabla^TH\cdot D\cdot \nabla H }{\dot{m}}.\ee
We thus obtain the expression for $f'(H)$ given in Eq.(10) of the main text.
%\be \label{eq:expression:f'}
%f'(H)=\frac{-\int_{m_1}^{m_2}dm\, \dot{m}(m, H)\, g(m, \dot{m}(m, H))}{\int_{m_1}^{m_2}\frac{dm}{\dot{m}} \,\nabla^TH\cdot D\cdot \nabla H}
%\ee
%with $\dot{m}(m, H)=\sqrt{2(H-V(m))}$ and  $\nabla^TH\cdot D\cdot \nabla H= D_{11}V'(m)^2+2D_{12}V'(m)\dot{m}(m, H)+D_{22}\dot{m}(m, H)^2$. 

Note that the effective potential $V(m)$ intervenes in Eq.~(8) [main text] in both the definition of the Hamiltonian $H$ and the function $f(H)$, through its derivative $f'(H)$ given by Eq.~(10)[main text]. Hence the functional forms of $H(m,\dot{m})$ and of $f(H)$ cannot be decoupled.

\subsection{Expression of $f(H)$ for particular cases}

\subsubsection{Elliptic limit cycle}

The continuous transition from a paramagnetic phase to an oscillating phase is well described using 
\be\begin{aligned} &V(m)=\frac{1}{2}v_0 m^2,\\ &g(m,\dot{m})=\alpha_0 \ve-\alpha_{1}m^2-\alpha_{2}m\dot{m}-\alpha_{3}\dot{m}^2\end{aligned}\ee 
where $\alpha_0$ is defined such that $\ve$ is a dimensionless parameter. 
We obtain from Eq.~(10) of the main text that $f(H)$ takes the generic form \be \label{eq:fH:elliptic_SM} f(H)=-\ve a H+bH^2 \ee 
where
\be \label{eq:expression:a}
a=\frac{\alpha_0}{D_{22}+D_{11}v_0}
\ee
and 
\be \label{eq:expression:b}
b=\frac{\alpha_{1}+3\alpha_{3}v_0}{4v_0(D_{22}+D_{11}v_0)}.
\ee 
The case $\ve<0$ corresponds to the time-independent paramagnetic phase ($H^*=0$), whereas $\ve>0$ corresponds to an oscillating phase with $H^*=\frac{\ve a}{2b}$.

\subsubsection{Non-elliptic limit cycle}

Close to a tricritical point where the paramagnetic, ferromagnetic and oscillating phase meet, $v_0$ changes sign. For $v_0=0$, we have  \be\begin{aligned} &V(m)=\frac{1}{4}v_1 m^4,\\ &g(m,\dot{m})=\alpha_0 \ve-\alpha_{1}m^2-\alpha_{2}m\dot{m}-\alpha_{3}\dot{m}^2.\end{aligned}\ee 
In that particular case, $f(H)$ takes the nonanalytic form 
\be \label{eq:fH:nonelliptic} f(H)=-\ve a H+cH^{3/2}\ee 
where
\be \label{eq:expression:c:2}
a=\frac{\alpha_0}{D_{22}}
\ee
and
\be \label{eq:expression:c}
c = \frac{8}{5\pi^2} \Gamma\left(\frac{3}{4}\right)^4 \frac{\alpha_1}{D_{22}\sqrt{v_1}}
\ee
%\be c=\frac{2a_1}{D_{22}\sqrt{v_1}}\frac{\Gamma(7/4)\Gamma(3/4)}{\Gamma(9/4)\Gamma(1/4)}\ee
where $\Gamma$ refers to the Euler Gamma function $\Gamma(x)=\int_0^{\infty}dt\,t^{x-1} e^{-t}$. The limit cycle for $H^*=\left(\frac{2\ve a}{3c}\right)^2$ has a non-elliptic form as $H=\frac{1}{4}v_1m^4+\frac{1}{2}\dot{m}^2$.
The scalings of $m$ and $\dot{m}$ with $\ve$ are different $m\sim\ve^{1/2}$ and $\dot{m}\sim \ve^{1/4}$.

\section{Overlap between spin configurations}
As the spins are exchangeable random variables in mean-field models, de Finetti's representation theorem \cite{hewitt_symmetric_1955_SM,aldous_ecole_1985_SM}
leads for large $N$ to
     \begin{equation} \label{eq:deFinetti}
        P(\{s_i\}) = \int_{-1}^1 \! \mathrm{d}m\, \tilde{P}(m)\, \mathcal{P}(\{s_i\}|m)
    \end{equation}
with a factorized conditional distribution $\mathcal{P}(\{s_i\}|m)$,
     \begin{equation}
       \mathcal{P}(\{s_i\}|m) =
       %\prod_{i=1}^N \frac{1}{2}\sqrt{1-m^2}\; \gamma(m)^{s_i},
       \left(\frac{1-m^2}{4}\right)^{N/2} \prod_{i=1}^N \left( \frac{1+m}{1-m} \right)^{s_i/2}
     \end{equation}
and $\tilde{P}(m)=\int d\dot{m}\, P(m, \dot{m})$.

To describe the overlap statistics, we introduce the probability distribution
$P(q)$ of the overlap $q$,
    \begin{equation}
      \label{eq:Poverlap}
        P(q)=\sum_{\{s_i^a\}, \{s_i^b\}} P(\{s_i^a\})P(\{s_i^b\}) \,\delta\bigg(\frac{1}{N}\sum_{i=1}^N s_i^as_i^b-q\bigg),
    \end{equation}
obtained by averaging over two statistically independent spin configurations $\{s_i^{a}\}$ and $\{s_i^{b}\}$.
%The probability density $P(\{s_i\})$ is given in Eq.~(11) of the main text.
%using de Finetti's representation theorem \cite{hewitt_symmetric_1955,aldous_ecole_1985}
%
Defining the Fourier transform (i.e., the characteristic function) $\chi(\omega)$ of the overlap distribution $P(q)$,
\be
\chi(\omega)=\int_{-1}^1 dq\, P(q)\, e^{i\omega q},
\ee
one finds after some algebra
\be
\chi(\omega)=\iint dm_a dm_b \, \tilde{P}(m_a)\, \tilde{P}(m_b)\, e^{i\omega m_am_b}\,.
\ee
Taking the inverse Fourier transform, one then obtains for the overlap distribution
\be     \label{eq:overlap_result}
P(q)=\iint dm_a dm_b \, \tilde{P}(m_a)\, \tilde{P}(m_b)\, \delta(m_am_b-q)\,.
\ee
For the paramagnetic phase ($\ve<0$), one has for $N\to\infty$,
\be \label{eq:p:m:para} \tilde{P}(m)=\delta(m)\,,\ee  
while for the elliptic limit cycle ($v_0>0$ and $\ve >0$), one instead finds,
\begin{equation}
  \label{eq:p:m:LC:elliptic}
  \tilde{P}(m) = \frac{1}{\pi} \, \left|\frac{ \ve a }{bv_0}-m^2 \right|^{-1/2}\,.
\end{equation}

%\begin{figure}[t]
%    \centering
%    \includegraphics{psi_y.pdf}
%    \caption{Scaling functions $\psi(y)$ for the elliptic limit cycle phase ($v_0>0$ and $\ve>0$) and $\tilde{\psi}(y)$ for the non-elliptic limit cycle phase ($v_0=0$ and $\ve>0$).}
%    \label{fig:overlap:psi_y}
%\end{figure}

Applying these results to the different phases and using Eqs.~\eqref{eq:overlap_result}, \eqref{eq:p:m:para} and \eqref{eq:p:m:LC:elliptic}, % and  \eqref{eq:p:m:LC:nonelliptic}
we obtain for the paramagnetic phase ($\ve <0$), $P(q)=\delta(q)$. 
For the elliptic limit cycle phase ($\ve>0$ and $v_0>0$), we have $P(q)=q_{\ve}^{-1} \psi(q/q_\ve)$, with $q_{\ve}=a\ve/bv_0$ and the scaling function $\psi(y)$ which is independent of $\ve$, 
\begin{equation} \label{eq:psi:scalfn}
      \psi(y) = \frac{2}{\pi^2} \int_{|y|}^{1} \frac{\mathrm{d}x}{\sqrt{(1-x^2)(x^2-y^2)}}\,\theta(1-|y|).
\end{equation}
We plot the scaling function in Fig.~1(a) of the main text.
%For the non-elliptic limit cycle phase ($\ve>0$ and $v_0=0$), we obtain the same form $P(q)=\tilde{q}_{\ve}^{-1} \psi_1(q/\tilde{q}_{\ve})$, with $\tilde{q}_{\ve}=4\ve a/3\sqrt{v_1}c$ and with a different scaling function $\tilde{\psi}(y)$
%\begin{equation} \label{eq:psi:scalfn:nonelliptic}
%      \tilde{\psi}(y) = 2d^2 \int_{|y|}^{1} \frac{\mathrm{d}x\, x}{\sqrt{(1-x^4)(x^4-y^4)}}\, \theta(1-|y|).
%\end{equation}
The scaling function $\psi(y)$ has a logarithmic divergence for $y \to 0$, and a non-zero limit $\psi(\pm1) = 1/\pi$
at the support boundaries.

\section{Specific spin model}

\subsection{Determination of the function $Y(m, \dot{m})$}
For the kinetic mean-field Ising model with ferromagnetic interactions given in the main text, one has %we have $\dot{m}=-m+\tanh[\beta(J_1m+h)]$, such that
\begin{align} \textbf{d}_{1}&=\left(-2, 2 -2\beta J_1+2\beta J_1(m+\dot{m})^2\right),\\
 \textbf{d}_{2}&=-\left(0, -2\beta +2\beta (m+\dot{m})^2\right),\end{align}
and 
\begin{align}
W_{1}^{\pm}&=\frac{1\pm m}{2}\big(1+\exp[\pm 2\beta (J_1m+h)]\big)^{-1},\\
W_{2}^{\pm}&=\frac{1\pm h}{2}\big(1+\exp[\pm 2\beta (J_2h+(1-\mu)m)]\big)^{-1},
\end{align}
where $h(m, \dot{m})=-J_1 m+\beta^{-1}\tanh^{-1}[m+\dot{m}]$. From Eq.~(6) of the main text, one can get the expressions of $Y(m, \dot{m})$, $V'(m)=-Y(m, 0)$ and of $D_{11}$, $D_{12}$ and $D_{22}$,
\begin{widetext}
\begin{align}
&\begin{aligned}   Y(m, \dot{m})=&\beta J_1m+(-1+\beta J_1)\dot{m}-\tanh^{-1}(m+\dot{m})+\beta \tanh[J_2\tanh^{-1}(m+\dot{m})+\beta(1-\mu-J_1J_2)m]\\&+(m+\dot{m})^2\left[\tanh^{-1}(m+\dot{m})-\beta\tanh[J_2\tanh^{-1}(m+\dot{m})+\beta(1-\mu-J_1J_2)m]-\beta J_1(m+\dot{m})\right],\end{aligned}\\
&V'(m)=-\beta J_1m +\beta J_1m^3+(1-m^2)\tanh^{-1}(m)-\beta(1-m^2)\tanh[J_2\tanh^{-1}(m)+\beta (1-\mu-J_1J_2)m],\\
& D_{11}=1-m\left(m+\dot{m}\right), \\
& D_{12}=\left(1-m(m+\dot{m})\right)\left(-1+\beta J_1-\beta J_1(m+\dot{m})^2\right),\\
& \begin{aligned}D_{22}=&\beta^2(1-(m+\dot{m})^2)\left[1-\left(\beta^{-1}\tanh^{-1}(m+\dot{m})-J_1m\right)\tanh\left(J_2\tanh^{-1}(m+\dot{m})+\beta(1-\mu-J_1J_2)m\right)\right]\\
&+\left(1-m(m+\dot{m})\right)\left(-1+\beta J_1-\beta J_1(m+\dot{m})^2\right)^2.\end{aligned}
\end{align}

\end{widetext}
%\be \begin{aligned}&(\dot{m}, Y(m, \dot{m}))=\frac{\textbf{a}_{+1}}{2}\dot{m}%\\&+\frac{\textbf{a}_{+2}}{2}(-h+\tanh[\beta(J_2h+(1-\mu)m)]).\end{aligned}\ee
%\textcolor{red}{ecrire $Y(m, 0)$ et $\dot{m}g(m, \dot{m})$. }

%As $Y(m, \dot{m})$ is analytic in $m$ and $\dot{m}$, $g(m, \dot{m})=\frac{Y(m, \dot{m})-Y(m, 0)}{\dot{m}}$ exists and is well defined. 

\subsection{Values of the different coefficients}
The coefficients ($\ve$, $a$, $b$ and $c$) of the large deviation function of Eq.~\eqref{eq:fH:elliptic_SM} and Eq.~\eqref{eq:fH:nonelliptic} are expressed in terms of the coefficients $v_0$ and $v_1$ of the series expansion of $V(m)$,
\be V(m)=\frac{v_0}{2}m^2+\frac{v_1}{4}m^4,\ee
of $\ve$, $\alpha_0$, $\alpha_1$ and $\alpha_3$ of the series expansion of $g(m, \dot{m})$,
\be g(m,\dot{m})=\alpha_0\ve-\alpha_{1}m^2-\alpha_{2}m\dot{m}-\alpha_{3}\dot{m}^2, \ee
and of $D_{11}(0,0)$ and $D_{22}(0,0)$, see Eqs.~\eqref{eq:expression:a}, \eqref{eq:expression:b} and \eqref{eq:expression:c}.

For the kinetic mean-field Ising model with ferromagnetic interactions given in the main text, all those coefficients can be expressed using the parameters $J_1$, $J_2$ controlling spin-spin or field-field interactions, $T$ the temperature and $\mu$ controlling the distance to equilibrium. 
We have the following relations, 
\begin{widetext}
\begin{align}
&\ve=(T_c-T)/T_c,\\
&\alpha_{0}=2 T_c/T,\\
& v_0=(\mu-1)/T^2+(1-J_1/T)(1- J_2/T),\\ 
&v_1=-2/3+(2J_2+3J_1)/3T-(\mu-1+J_1J_2)/T^2-(\mu-1-J_2T+J_1J_2)^3/3T^4, \\
&\alpha_{1}=-2 + (2 J_2 + J_2^3 + 3 J_1)/T -  2(1 + J_2^2) (-1 + J_1 J_2 +\mu)/T^2+ J_2 (-1 + J_1 J_2 +\mu)^2/T^3,\\
&\alpha_{3}=-2/3 +(2J_2 + J_2^3 + 3J_1)/3T,\\
& D_{11}(0,0)=1,\\
& D_{22}(0,0)=1/T^2+(J_1/T-1)^2.
\end{align}
\end{widetext}

\subsection{Sections of $\phi(m, \dot{m})$}
We plot sections at constant $m$ or $\dot{m}$ values of $\phi(m, \dot{m})$ showed in Fig.~2 of the main text when both the paramagnetic and the oscillating phase are locally stable. In Fig.~1(a) and 1(b) we plot a section at $m=0$ and in Fig.~1(c) and 1(d) a section at $\dot{m}=0$.
\begin{figure}[htbp!]
    \centering
    \includegraphics{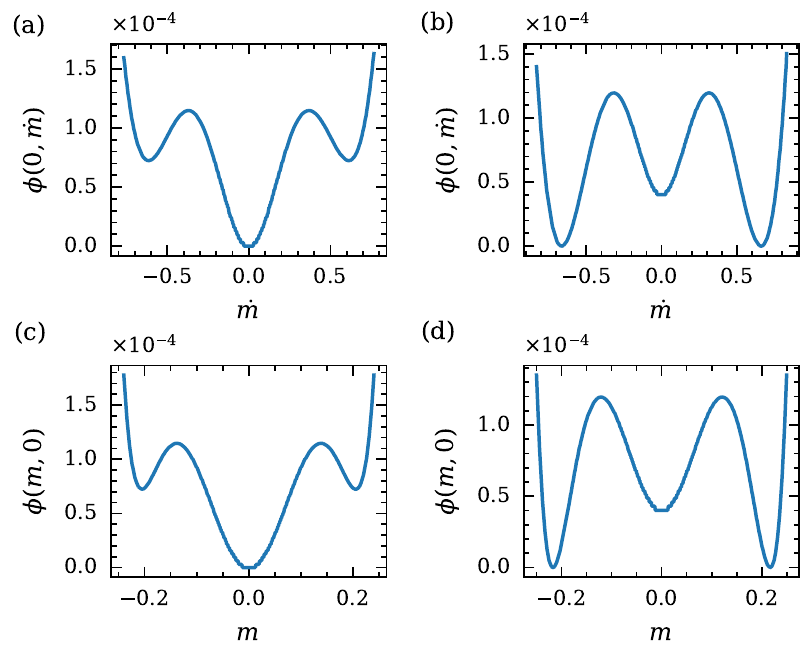}
    \caption{Sections of $\phi(m, \dot{m})$ at $m=0$ [(a) and (b)] and at $\dot{m}=0$ [(c) and (d)]. (a), (c) correspond to Fig.~2(c) of the main text and (b), (d) to Fig.~2(d) of the main text. }
    \label{fig:SM:phi:cut}
\end{figure}

\section{Entropy production}
The transition to a limit cycle may also be characterized thermodynamically as a transition from microscopic to macroscopic irreversibility, by introducing the entropy production density $\sigma=\Sigma/N$ in the limit $N\to\infty$, where the steady-state entropy production $\Sigma$ identifies with the entropy flux \cite{schnakenberg_1976_SM,gaspard_time-reversed_2004_SM},
\be
\label{eq:def:entroy:prod}
\Sigma = \frac{1}{2} \sum_{\mC,\mC'} \big[ W(\mC'|\mC)P(\mC)-W(\mC|\mC')P(\mC')\big]\, \ln \frac{W(\mC'|\mC)}{W(\mC|\mC')} \,.
\ee
One finds (see below) that in the paramagnetic phase ($\epsilon<0$), $\sigma=0$ while in the oscillating phase ($\epsilon>0$), $\sigma \sim \epsilon$ becomes non-zero
(similar calculations have been performed in \cite{xiao_entropy_2008_SM,seara_irreversibility_2021_SM} in the context of chemical oscillators).
Hence the entropy production density $\sigma$ is also an order parameter of the phase transition to a limit cycle, associated with a macroscopic breaking of time-reversal invariance. The corresponding critical exponent, equal to $1$, is
the same as for the order parameter $\langle \dot{m}^2\rangle$ characterizing the spontaneous breaking of time translation invariance.
In the nonequilibrium paramagnetic or ferromagnetic phases, the entropy production $\Sigma$ remains microscopic, i.e., $\Sigma = O(N^0)$.
Note that the transition between paramagnetic and ferromagnetic nonequilibrium phases is expected to be characterized by a cusp of the entropy production $\Sigma$ \cite{noa_entropy_2019_SM}.
We evaluate here the entropy production in spin models under the assumptions detailed in Sec.~\ref{sec:trans:rate}.
We show that we recover results obtained in the diffusive limit as done in \cite{xiao_entropy_2008_SM,seara_irreversibility_2021_SM} in the context of chemical reactions.

\subsection{Spin-reversal dynamics}
We consider that there are only two different types of transitions $k=a, b$.
Expressing Eq.~\eqref{eq:def:entroy:prod} in the variables $m$ and $\dot{m}$ and considering the lowest order in N, the entropy production density $\sigma=\Sigma/N$ becomes
\be \sigma=\sum_{k} \left\langle\left(W_k^+(m, \dot{m})-W_{k}^-(m, \dot{m})\right) \ln\frac{W_k^+(m, \dot{m})}{W_{k}^-(m, \dot{m})}\right\rangle.
\ee
%\be
%\sigma=\iint dm d\dot{m}P(m, \dot{m})\sum_{k>0}\frac{\left(W_k(m, \dot{m})-W_{-k}(m, \dot{m})\right)^2}{W_k(m, \dot{m})}
%\ee
We assume for now that $1-W_{k}^-/W_k^+$ is small, such that the entropy production can be approximated as
\be \label{eq:approx:sigma}
\sigma=\left\la\, \sum_{k}\frac{\left(\, W_k^+(m, \dot{m})-W_{k}^-(m, \dot{m})\, \right)^2}{W_k^+(m, \dot{m})}\, \right\ra.
\ee
We note $\mathbf{A}$ the change-of-basis matrix such that $\mathbf{A}\mathbf{d}_1=(1, \,0)$ %\begin{pmatrix}1\\0\end{pmatrix}$
and $\mathbf{A}\mathbf{d}_2=(0, \,1)$. %\begin{pmatrix}0\\1\end{pmatrix}$. 
Using that $(\dot{m},\, Y(m, \dot{m}))=\sum_{k}\left(W_k^+-W_{k}^-\right) \mathbf{d}_{k}$  and that $\mathbf{D}=\sum_{k}W_k^+\mathbf{d}_{k}\!\cdot\!\mathbf{d}_{k}^T$ at first order in $1-W_{k}^-/W_k^+$, in the new basis we have
\be\label{eq:SM:mdot:Y:Wk} \begin{pmatrix}W_1^+(m, \dot{m})-W_{1}^-(m, \dot{m})\\W_2^+(m, \dot{m})-W_{2}^-(m, \dot{m}) \end{pmatrix}=\mathbf{A} \begin{pmatrix} \dot{m}\\Y(m, \dot{m})\end{pmatrix}\ee
%\be\begin{pmatrix} \dot{m}\\Y(m, \dot{m})\end{pmatrix}=A^{-1}\cdot \begin{pmatrix}W_1(m, \dot{m})-W_{-1}(m, \dot{m})\\W_2(m, \dot{m})-W_{-2}(m, \dot{m}) \end{pmatrix}\ee
and 
\be  \textbf{Diag}\left(W_1^+(m, \dot{m}), W_2^+(m, \dot{m})\right)=\mathbf{A}\cdot\mathbf{D}\cdot \mathbf{A}^T.\ee 
%\be D=A^{-1}\cdot \text{Diag}\left(W_1(m, \dot{m}), W_2(m, \dot{m})\right)\cdot %\begin{pmatrix}W_1(m, \dot{m}) &0\\0& W_2(m, \dot{m})\end{pmatrix}
%(A^T)^{-1}.\ee 
Hence, the entropy production density can be rewritten as 
\be \label{eq:entropy:prod:SM1}
\sigma= \big\la \, (\dot{m}, Y(m, \dot{m}))^T\cdot\mathbf{D}^{-1}\cdot  (\dot{m}, Y(m, \dot{m}))\, \big\ra.
\ee
Retaining the lowest order of $Y(m, \dot{m})$ in $\ve$ and using that $\langle V'(m)\dot{m}\rangle=0$ when $\phi(m, \dot{m})=f(H)$ with $H=V(m)+\frac{\dot{m}^2}{2}$, the entropy production density becomes
\be
\sigma=\big\la \,(D^{-1})_{11}\dot{m}^2+(D^{-1})_{22}V'(m)^2\, \big\ra.
\ee
In the paramagnetic phase ($\ve<0$), one has for finite $N$ the scaling $\sigma \sim N^{-1}$ because, as shown in the main text,
\be
\la \dot{m}^2\ra\sim N^{-1},\qquad  \la V'(m)^2\ra \sim N^{-1}.
\ee
Therefore in the limit $N\to\infty$, the entropy production density vanishes.
In the oscillating phase ($\ve>0$), the entropy production density $\sigma$ converges to a finite value $\sigma\sim \ve$ when $N\to\infty$, due to the fact that 
\be
\la \dot{m}^2\ra\sim \ve,\qquad  \la V'(m)^2\ra \sim \ve.
\ee

To obtain these results for the entropy production, we assumed that  $1-W_k^-(m, \dot{m})/W_k^+(m, \dot{m})$ was small. Under the assumption that $W_k^{\sigma}$ is of order $1$ (which is verified in the specific model on spins and fields), it is equivalent to assuming that $W_k^+-W_k^-$ is small. 
One has that $W_k^{+}(m, \dot{m})-W_k^-(m, \dot{m})=y^k$ where
\begin{equation}
\textbf{y}=(y^a, y^b)=\mathbf{A}\begin{pmatrix} \dot{m}\\ Y(m, \dot{m}) \end{pmatrix}
\end{equation}
[see Eq.~(\ref{eq:SM:mdot:Y:Wk})], and writing
\begin{equation}
\textbf{x}=\mathbf{A}\begin{pmatrix} m\\ \dot{m} \end{pmatrix}
\end{equation}
from Eqs.~(\ref{eq:SM:dt:av:m:is:mdot}) and (\ref{eq:dmdot:dt:average:SM}) one gets that
\be \frac{d\langle \textbf{x}\rangle }{dt}=\langle \textbf{y}\rangle. \ee
In a static phase $\langle \textbf{y}\rangle=0$ and thus we expect this value to remain small close to the transition line for continuous phase transitions. Therefore, the assumption that $W_k^+-W_k^-$ is small is valid close to transition lines, for continuous transition from a static phase to either another static phase or to an oscillating phase. 
%, we have that $W_{1}(m,\dot{m})-W_{-1}(m, \dot{m})=\dot{m}/2$ and $W_{2}(m, \dot{m})-W_{-2}(m, \dot{m})=\dot{h}/2$ with $\dot{h}$ defined in Eq.~\ref{eq:dyn:h}. At the transition between the paramagnetic phase to an elliptic limit cycle phase, $\dot{m}, \dot{h}=O(\varepsilon)$, and $W_1, W_2$ are of order $1$, hence the previous assumption is verified.

\subsection{Generic diffusive limit}
The entropy production density may also be derived in the diffusive limit, as done by \cite{xiao_entropy_2008_SM,seara_irreversibility_2021_SM}.
The linear and quadratic terms in $\nabla \phi$ of Eq.~(3) of the main text correspond to a Fokker-Planck equation on $P(m, \dot{m},t)$
\be \label{eq:fokker:planck}
\partial_{t} P = \partial_m(\dot{m}P)+\partial_{\dot{m}}Y(m, \dot{m})P+\frac{1}{N}\sum_{i,j} \partial_i\partial_j (D_{ij} P).
\ee
We write  $\mathbf{x}=(m, \dot{m})$ and $\mathbf{y}=(\dot{m}, Y(m, \dot{m}))$.  We introduce the probability current density $J(\mathbf{x})$ such that Eq.~\eqref{eq:fokker:planck} becomes $\partial_t P=-\nabla\cdot \mathbf{J}$ with at the lowest order in $N$
\be \mathbf{J}(\mathbf{x}, t)=-\left(\mathbf{y}+\frac{1}{N}\mathbf{D}\cdot \nabla \right) P(\mathbf{x}, t) +O(N^{-1}).\ee

We consider a trajectory $\mathbf{x}(t)=(m(t), \dot{m}(t))$.  We define an entropy along the trajectory, as done by \cite{seifert_entropy_2005_SM}, 
\be s(t)=-\ln P(\mathbf{x}(t), t).\ee
The rate of change of the entropy along a trajectory is
\be \dot{s}=-\frac{\partial_t P(\mathbf{x}, t)}{P(\mathbf{x}, t)} -\frac{1}{P(\mathbf{x}, t)}\, \mathbf{\dot{x}}^T\cdot\nabla P(\mathbf{x}, t)\,.\ee
Using the probability current density $\mathbf{J}$, we obtain 
\be \label{eq:sdot:traj}
\dot{s}=\left[-\frac{\partial_t P}{P} +\frac{2N}{P}\mathbf{\dot{x}}^T \cdot \mathbf{D}^{-1} \cdot  \mathbf{J}\right]+N\, \mathbf{\dot{x}}^T \cdot \mathbf{D}^{-1} \cdot  \mathbf{y}.
\ee
The term within the square bracket in Eq.~\eqref{eq:sdot:traj} denotes the trajectory-dependent total entropy production $\dot{s}_{tot}$ as shown in \cite{seifert_entropy_2005_SM}.
We define the medium entropy production as
\be \dot{s}_m=N\,\mathbf{\dot{x}}^T \cdot \mathbf{D}^{-1}\cdot \mathbf{y}. \ee
In the limit $N\to\infty$, $\dot{\mathbf{x}}(t)=(\dot{m}, Y(m, \dot{m}))=\mathbf{y}$. Averaging over the stationary distribution, we obtain an expression for the entropy production density
\be \sigma=\frac{1}{N}\langle \dot{s}_m\rangle=\langle \mathbf{y}^T\cdot \mathbf{D}^{-1}\cdot \mathbf{y}\rangle \ee
which is consistent with the expression Eq.~\eqref{eq:entropy:prod:SM1} derived with the spin-reversal dynamics, under the approximation Eq.~\eqref{eq:approx:sigma}.

\end{document}